\documentclass[aps,preprint,dvipsnames]{revtex4-2}
\usepackage[french,english]{babel}
\usepackage{amsmath,amssymb}
\usepackage{graphicx,fourier}
\usepackage{verbatim}
\usepackage{color}
\usepackage{hyperref}
\usepackage{eqnarray,tabularx}
\usepackage{ccaption,float,esvect,amsthm,picins}
\usepackage{wrapfig}
\usepackage{amsbsy,amssymb,bm}
\usepackage{pstricks,pst-circ,pstricks-add,pst-node,pst-coil,pst-eucl}
\usepackage{amsmath}
\usepackage{subfigure,xcolor}

\begin{document}

\title{ Resonance curves are perfect circles}
\author{\textcolor{blue}{Adel H. Alameh}}
\affiliation{\textcolor{red}{Lebanese University, Department of Physics, Hadath, Beirut, Lebanon}}
\email{adel.alameh@eastwoodcollege}
\begin{abstract}
The ability to approach a physical phenomenon and grasp its major importance is a remarkable quality of understanding.  This paper presents a rather elegant and novel way of looking at the resonance phenomenon, which among others shares a common conceptual basis in various fields of physics. For the sake of simplicity, the discussion will be restricted to the case of electric current resonance in series $RLC$ circuits.\\

 The mathematics of electric resonance is thus meticulously extended to the extent that it ultimately unravels  an invaluable relationship between the current at a certain driving frequency and that at the resonance frequency. Much further it gives rise to an elaborate correlation between any two driving frequencies that give the same current.\\

Arising from the previously mentioned relations, a new technique is devised, a simple geometrical construction, that without running into tedious calculations, allows the computation of the  phase difference between the current and the impressed voltage at any given frequency. Not to mention another geometric utility, that is miraculously   constructed  to correlate any frequency with its corresponding ``conjugate'' one, that allows the same current in the circuit.

\end{abstract}
\begin{minipage}{\textwidth} \renewcommand{\textheight}{2\textheight}
\maketitle
\end{minipage}
\bigskip
\newpage
\section*{Response of an $RLC$ component to a sinusoidal signal}
\noindent Consider an $RLC$ series circuit and a generator $G$ of negligible resistance.
\, At the instant $t_0=0$, we close the switch. The voltage impressed by $G$ is alternating sinusoidal of the form \mbox{$u_g(t)=U_m\sin(\omega t +\varphi) $}.
It is required to study the response of the circuit to such a signal.\\
\noindent Before delving into details, it is essential to highlight the following primary issues$\colon$ \vspace{-.2cm}
\begin{itemize}
\item[$\bullet$] The voltage $u_R$ across a resistor is the image of the current $i$ to a multiplicative constant $R$\,. It follows that a sinusoidal voltage across a resistor allows a sinusoidal current of same angular frequency and same initial phase angle.
\item[$\bullet$] The voltage $u_L$ across a coil of inductance $L$ and negligible resistance is the image of $\displaystyle\frac{di}{dt}$ to a multiplicative constant $L$. Accordingly a sinusoidal voltage across an inductor allows a sinusoidal current of same angular frequency and lagging behind the voltage by a phase angle of $\displaystyle\frac{\pi}{2}$.

     \item[$\bullet$] The voltage $u_C$ across   a capacitor  $C$ is the image of $\int_{t_0}^t idt$ to a multiplicative constant $\displaystyle\frac{1}{C}$ \,. Hence a sinusoidal voltage across a capacitor allows a sinusoidal current of same angular frequency and leading the voltage by a phase angle of $\displaystyle\frac{\pi}{2}$\,.
 \end{itemize}
 \vspace{-.2cm}
  The case being so, we proceed by applying the law of addition of voltages$\colon$
 \begin{equation}u_R+u_L+u_C=u_G \label{lawofvoltages}\end{equation}
\noindent  we then plug into equation~(\ref{lawofvoltages}), the expressions $u_R=Ri$, $u_L=L\displaystyle\frac{di}{dt}$, and $u_C=\displaystyle\frac{\int_0^t idt}{C}$
  \begin{equation} Ri+L\displaystyle\frac{di}{dt}+\displaystyle\frac{\int_0^t idt}{C}=u(t) \label{explicit}\end{equation}
 \emph{ A lot of focus  is put in this paper exclusively on the steady state solution of equation~(\ref{explicit}).}
  The  voltage across the $RLC$ component is the sum of three sinusoidal voltages, hence as confirmed by the previously highlighted arguments, it would  be traversed by a current of the form \mbox{$i=I_0\sin\omega t$.}
Thus, in the explicit form equation~(\ref{explicit}) becomes
\begin{equation}RI_0\sin\omega t+L\omega I_0 \sin\left(\omega t+\displaystyle\frac{\pi}{2}\right) + \displaystyle\frac{I_0}{C\omega}\sin\left(\omega t-\displaystyle\frac{\pi}{2}\right)=U_m\sin(\omega t +\varphi) ~\label{explicit1} \end{equation} \vskip -.2cm
Instead of having recourse to trigonometric manipulations to deal with equation~(\ref{explicit1}), it is rather possible to resort to the very efficient technique devised by Fresnel deliberately for such purposes. It is about phasor diagrams which are a graphical way of representing  sinusoidal waveforms by phasor vectors.  Hence~\cite{zemansky}~\cite{Jcassa} .\\




\begin{equation}I_0=\displaystyle\frac{U_m}{\sqrt{R^2+\left(L\omega -\displaystyle\frac{ 1}{C\omega}\right)^2}~\label{intensity}}
 \end{equation}
 and
 \begin{equation}\tan\varphi=\displaystyle\frac{L\omega -\displaystyle\frac{1}{C\omega}}{R}~\label{phase}\end{equation}
 and naturally
 \begin{equation}\cos\varphi=\displaystyle\frac{R}{\sqrt{R^2+\left(L\omega -\displaystyle\frac{ 1}{C\omega}\right)^2}}~\label{cosinephi}\end{equation}
\section*{Variation of the peak current with the angular frequency}
\noindent We now turn our attention to study the variation of the peak value of the current $I_0$ in terms of the driving angular frequency $\omega$\,. It may therefore be expedient to limit our discussion to examine graphically the function $I_0=f(\omega)$ of equation~(\ref{intensity}). Figure (1) is a plot of $I_0$ in terms of $\omega$ for given fixed values of $R$\,, $L$\,, and $C$\,.
\begin{center}
\includegraphics[]{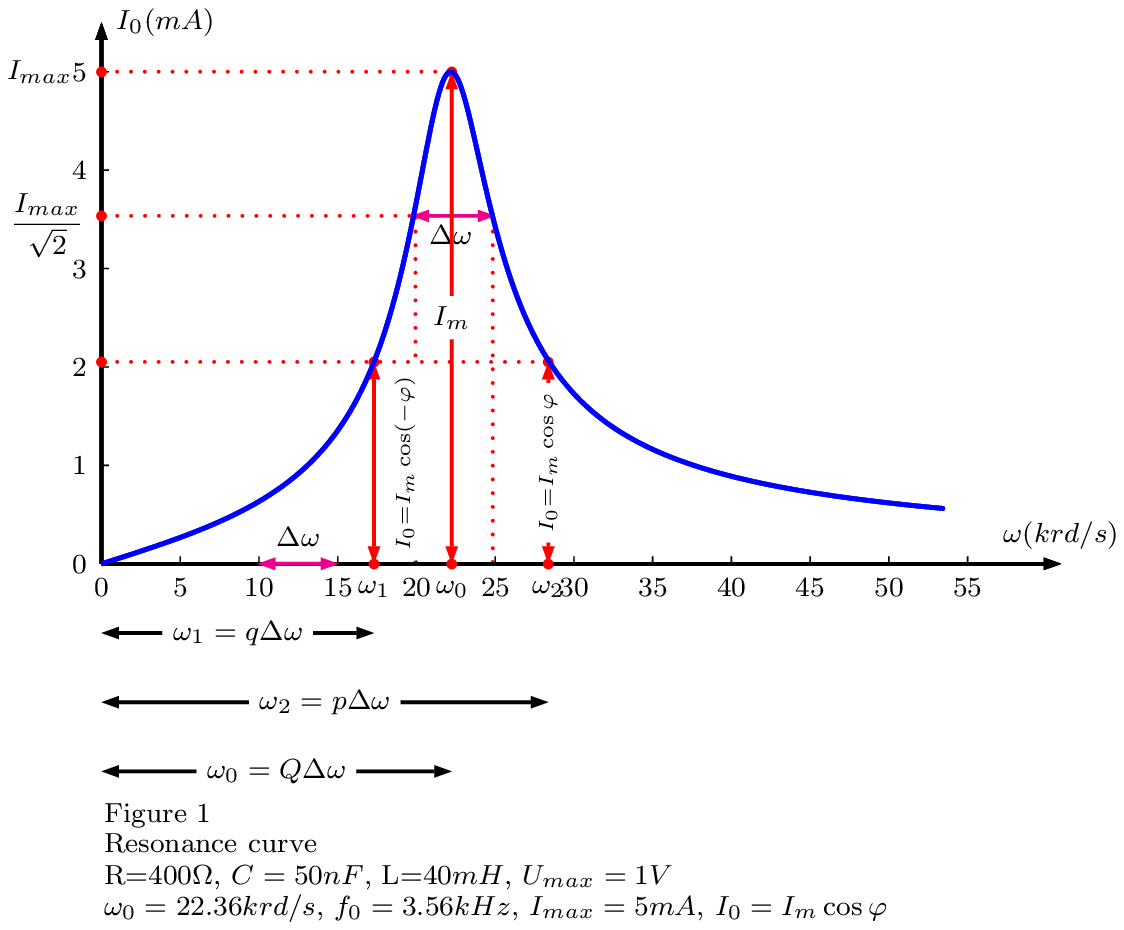}
\end{center}
Strictly speaking, there is a very significant case that deserves special mention for  its practical uses in communication equipments.  it is about having $L\omega =\displaystyle\frac{1}{C\omega}$\,, in this case the circuit is resistive, the circuit thus allows the passage of the highest current \begin{equation}I_{max}=\displaystyle\frac{U_m}{R}~\label{Imax}\end{equation}
 and $\tan\varphi$ given in equation~(\ref{phase}) reduces to zero. Hence the applied voltage and the current in the circuit are inphase. We then say that the circuit enters in current resonance. In other words, resonance occurs when the driving frequency $\omega$ is equal to $\omega_0$ where   $\omega_0=\displaystyle\frac{1}{\sqrt{LC}}$ is the natural or proper angular frequency of the circuit\,.
 However for frequencies comprised between $0$ and $\omega_0$ we have $\displaystyle\frac{1}{C\omega}>L\omega$\,, so the circuit is capacitive since $\tan\varphi$ is negative. Hence the  voltage across the circuit lags behind the current by the phase angle  $\varphi$\,. Moreover for frequencies lying in the interval ] $\omega_0$, $+\infty$[ we have $L\omega>\displaystyle\frac{1}{C\omega}$\,, so the circuit is inductive and $\tan\varphi$ is positive. Accordingly the voltage applied across the $RLC$ component leads the current by $\varphi$.

 An insightful aspect of equation~(\ref{intensity})   is  achieved after a little manoeuvre is brought about. For that reason the numerator and the denominator of the right side  of equation~(\ref{intensity}) are multiplied by $R$\,. Accordingly
 \begin{equation}I_0=\displaystyle\frac{U_m}{R}\cdot\displaystyle\frac{R}{\sqrt{R^2+\left(L\omega -\displaystyle\frac{ 1}{C\omega}\right)^2}~\label{intensity1}}\end{equation}
and using equations~(\ref{cosinephi}) and~(\ref{Imax}) gives
\begin{equation}I_0=I_{max}\cos\varphi ~\label{astounding} \end{equation}
Equation~(\ref{astounding}) is an astounding revelation, for it implies the circular character of  resonance curves. This impressive fact, is schematically illustrated  by figures (2a) and (2b) that show a plot of $I_0$ versus $\varphi$ in the polar system of coordinates.\\

\noindent\includegraphics[]{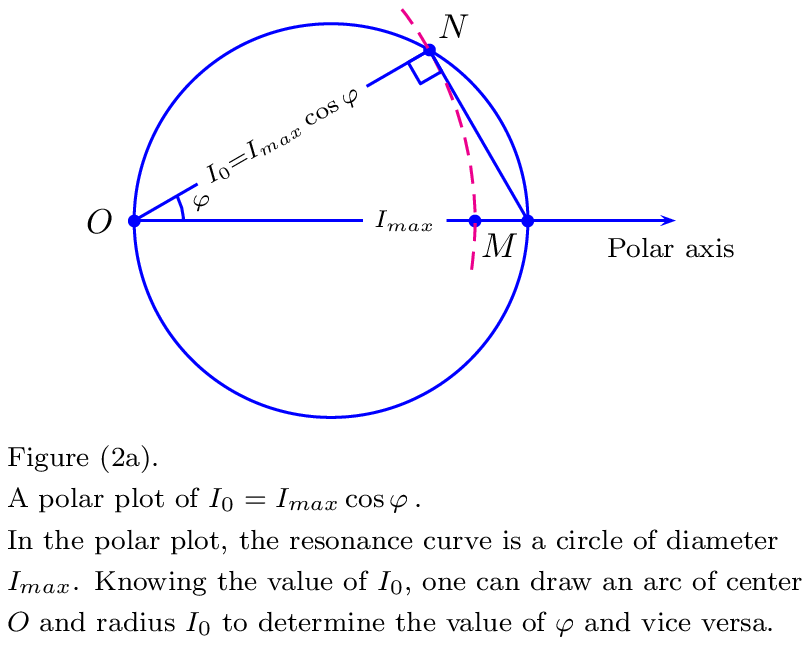}\hfill \includegraphics[]{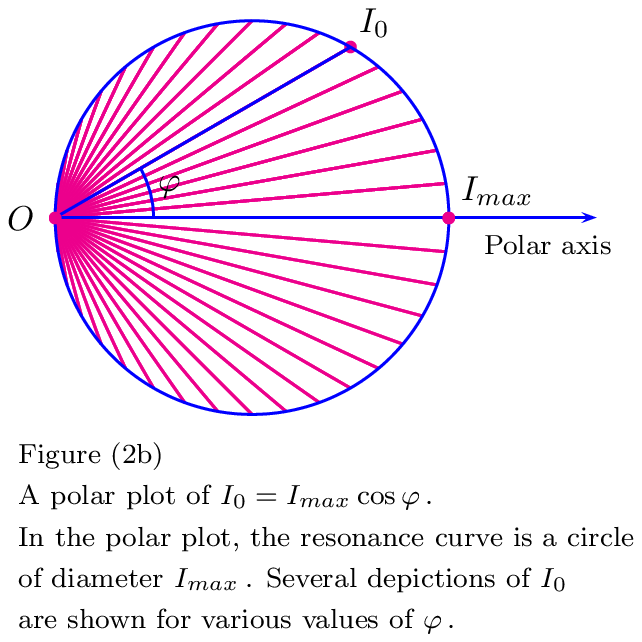}

\noindent Figure (2a) is a polar plot of $I_0=I_{max}\cos\varphi$\,, evidently $I_0$ traces a circle \cite{thomas} of diameter $I_{max}$. Given the value of $I_0$, one can draw an arc of center $O$ and radius $I_0$ to meet the resonance circle. The angle subtending the obtained  arc $\widearc{MN}$ is $\varphi$. Figure (2b) shows numerous values of $I_0$ for various values of $\varphi$.

\section*{Resonance and bandwidth}
\noindent In resonance instances, the resonance frequency $\omega_0$ is that at which the power transfer from the generator to the electric oscillator is maximum~\cite{Resnik}. There is, however, another range of frequencies of practical importance. Literally it is about the response band of frequencies surrounding $\omega_0$\,, and whose  extreme frequency values  allow  a transfer of half the maximum power delivered by the generator.  Bearing in mind the dependence of the power on $I_{eff}^2$ ~\cite{zemansky}\,,  we seek to determine the bounds of the response band by having the current to fall to $\displaystyle\frac{\sqrt{2}}{2}$ of its maximum value. Then, on the basis of equation~(\ref{intensity}) one needs to have
\begin{equation}R=\left|L\omega- \displaystyle\frac{1}{C\omega}\right|~\label{response}\end{equation}
There are two accepted values of $\omega$ that stem from equation~(\ref{response}), Namely \begin{equation}\omega_1=\displaystyle\frac{R}{L}\left(-\displaystyle\frac{1}{2}+\displaystyle\frac{1}{2}\sqrt{1+4\displaystyle\frac{L}{R^2 C}}\,\right)~\label{root1}\end{equation}
and
\begin{equation}\omega_2=\displaystyle\frac{R}{L}\left(\displaystyle\frac{1}{2}+\displaystyle\frac{1}{2}\sqrt{1+4\displaystyle\frac{L}{R^2 C}}\,\right)~\label{root2}\end{equation}
Definitely, one can subtract equation~(\ref{root1}) from equation~(\ref{root2}) to obtain the width of the response band
\begin{equation}\Delta \omega=\displaystyle\frac{R}{L}~\label{width}\end{equation}
It is as well interesting in this respect to note that
\begin{equation}\omega_1 \omega_2=\omega_0^2 ~\label{inverse1} \end{equation}
Certainly, relation~(\ref{inverse1}) is  widely known by people in the field; however it shall be later pursued at greater length to prove its validity for any couple of frequencies that give the same current.\\
\section*{Quality factor}
\noindent   In this section a good deal of attention shall be given to what is called the quality factor $Q$ of an electric oscillator.  In a general sense, the quality factor characterizes the sharpness of the resonance phenomenon~\cite{foundations}~\cite{tout} . It is defined as the ratio of $\omega_0$ to $\Delta\omega$
\begin{equation} Q=\displaystyle\frac{\omega_0}{\Delta \omega}=\displaystyle\frac{1}{R}\sqrt{\displaystyle\frac{L}{C}}~\label{qualityfactor}\end{equation}

 Despite the numerous definitions given to this factor, still there is room to introduce a brand new one that unveils  interesting consequences. Hence, I suggest to divide the frequency axis in figure (3) into regular parts each of width $\Delta \omega$\,. The case being so,  it turns out that \emph{ the quality factor $Q$ is the number of bandwidths $\Delta \omega$ included in $\omega_0$}. Following this reasoning, I define two new factors $q$ and $p$ which I name \emph{ pseudo quality factors} so that, any frequency $\omega_1$ in the interval $[0,\omega_0]$\,, can be expressed as
 \begin{equation}\omega_1=q \Delta\omega=q\left(\displaystyle\frac{R}{L}\right) \label{omega3}\end{equation}
 where $q$ is the number of bandwidths $\Delta\omega$ comprised in $\omega_1$\,, naturally  $0\leq q \leq Q$\,. Likewise, any frequency $\omega_2$ in the interval $[\omega_0,\infty[$\,, may be written as
 \begin{equation}\omega_2=p\Delta\omega = p\left(\displaystyle\frac{R}{L}\right)\label{omega4}\end{equation}
  where $p$ is the number of bandwidths $\Delta \omega$ embraced in $\omega_2$ and evidently  $p\geq Q$\,. It is to be noted here that relations~(\ref{omega3}) and~(\ref{omega4}) were inspired by relations~(\ref{root1}) and (\ref{root2}) and shall be reasonably justified in the process. Accordingly we intend to find a relation between a certain driving frequency $\omega_1$ and its corresponding conjugate $\omega_2$ that gives the same value of $I_0$\,. Curiously if we substitute for $\omega_1$ the value $q\left(\displaystyle\frac{R}{L}\right)$ and for $\omega_2$ the value $\displaystyle\frac{1}{qRC}$ successively in equation~(\ref{intensity}) we get
\begin{equation}I_0=\displaystyle\frac{U_m}{\sqrt{R^2+\left(qR-\displaystyle\frac{L}{qRC}\right)^2}}~\label{qRL}\end{equation}
and
\begin{equation}I_0=\displaystyle\frac{U_m}{\sqrt{R^2+\left(\displaystyle\frac{L}{qRC}-qR\right)^2}}~\label{qRC}\end{equation}
Surprisingly the values of $I_0$ given by equations~(\ref{qRL}) and~(\ref{qRC}) are equal, but the value  $\displaystyle\frac{1}{qRC}$ of $\omega_2$ must be equal to $p\left(\displaystyle\frac{R}{L}\right)$ as previously suggested by equation~(\ref{omega4}). This implies that
\begin{equation}pq=Q^2~\label{Qsquared}\end{equation}
 which can be written in the form \begin{equation}\displaystyle\frac{q}{Q}=\displaystyle\frac{Q}{p}~\label{ratio}\end{equation}
 Let's now rewrite the inequality $0\leq q\leq Q$\,, mentioned earlier in this section in the form
 \begin{equation} 0\leq \displaystyle\frac{q}{Q}\leq 1 ~\label{ratio1}\end{equation}
 This suggests to let the ratio $\displaystyle\frac{q}{Q}$ be equal to the cosine of an angle $\beta$ that is  envisaged for a special geometric  purpose to be illustrated later. Thus
 \begin{equation}\cos\beta=\displaystyle\frac{q}{Q}~\label{cosine1}\end{equation}
 and one can deduce from equation~(\ref{ratio}) that
   \begin{equation}p=\displaystyle\frac{Q}{\cos\beta}\label{cosine2}\end{equation}
By manipulating equations~(\ref{cosine1}) and~(\ref{cosine2}) with equations~(\ref{qualityfactor}),~(\ref{omega3}), and~(\ref{omega4}) we get
\begin{equation}\omega_1=\omega_0 \cos\beta \label{cosine5}\end{equation}
and
\begin{equation}\omega_2=\displaystyle\frac{\omega_0}{\cos\beta}\label{cosine6}\end{equation}
furthermore the multiplication of equations~(\ref{cosine5}) and (\ref{cosine6}) yields
\begin{equation}\omega_1 \omega_2 =\omega_0^2\label{harmonic}\end{equation}
It becomes increasingly apparent that any pair of frequencies $\omega_1$ and $\omega_2$ that allow the same current in the circuit are tied by relation~(\ref{harmonic}). As such, and in the frequency space, $\omega_2$ is mathematically~\cite{wiki} identified  to be the inverse of $\omega_1$ with respect to a reference circle of radius $\omega_0$ and  center the frequency $0$ .  This fact is explicitly displayed in figure (3). It is evident that the point $C(\omega_2)$ is the harmonic conjugate of $A(\omega_1)$ with respect to the points $S(-\omega_0)$ and $T(\omega_0)$ as inferred from relation~(\ref{harmonic}). In other words, the four points C, A, S, and T form a harmonic tetrad in Galois geometry~\cite{galois}.\vspace{-.75cm}
\begin{center}
\includegraphics[]{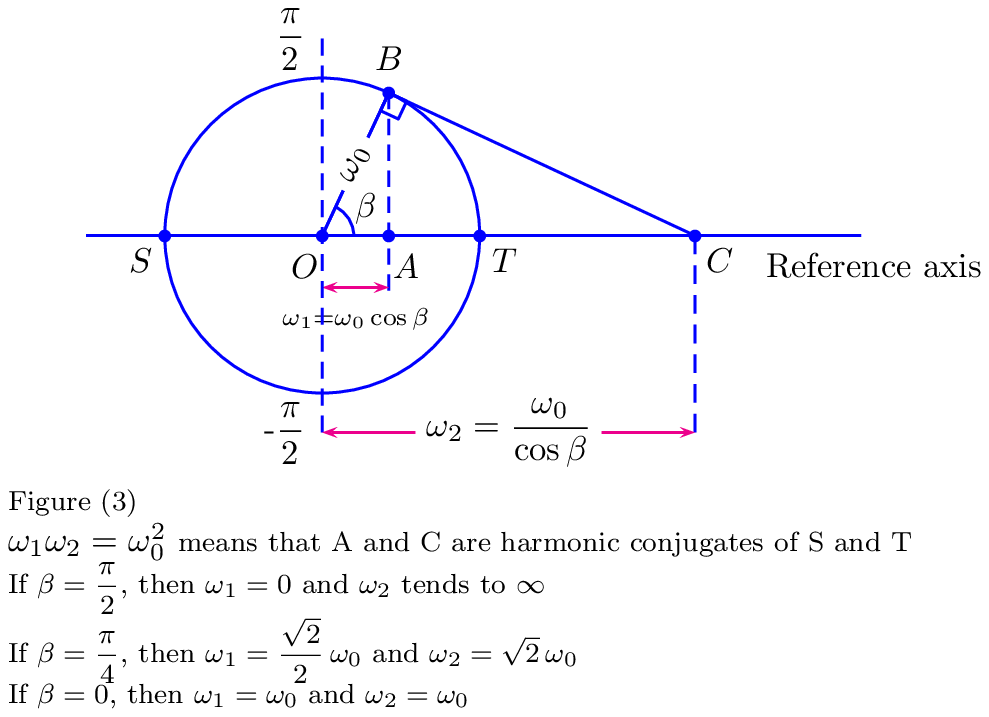}
\end{center}\vspace{-1.1cm}
\section*{Conclusion} \vspace{-.5cm}
\noindent The traditional pattern of the resonance curve is schematically displayed in figure (4). The relation $I_0=I_{max}\cos\varphi$ reveals an invaluable asset in sofar as it enables graphically the computation of the phase difference $\varphi$ at any given frequency. Additionally it provides by its circular character in the polar system an elaborate tool to compute the phase difference  $\varphi$ given the values of $I_0$ and $I_{max}$\,. Dividing the frequency axis in equal portions of $\Delta \omega$ allows to find prominent relations between any pair of conjugate frequencies and the resonant frequency $\omega_0$\,. Not to mention the special attention given to the relation $\omega_1 \omega_2 =\omega_0^2$ that ties any pair of conjugate frequencies giving the same value of $I_0$\,.\vspace{-.12cm}
\begin{center}
 \includegraphics[height=9cm]{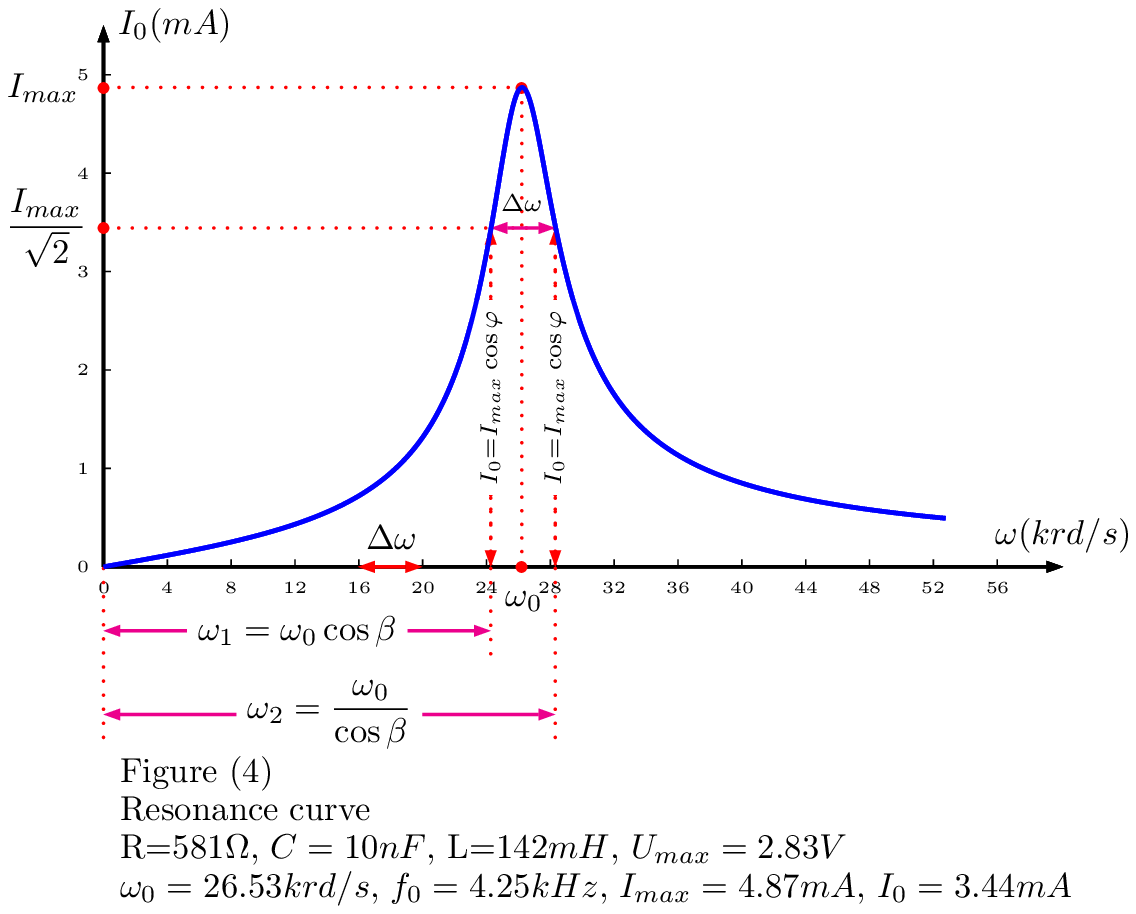}
\end{center}


\vspace{-1cm}
\begin{thebibliography}{5}
\bibitem{zemansky}Sears and Zemansky. \emph{University Physics}, Thirteenth Edition (Addison-Wesly, Boston), p. 1030.
\bibitem{Jcassa}J.Cessac, Georges Tr\'eherne. \emph{Physique, classe terminale C},1966 Paris, p. 242.
\bibitem{thomas}George B.Thomas. \emph{Thomas' Calculus}, Tenth Edition (Addison-Wesly, Boston), p.765.
\bibitem{Resnik}Resnick and Halliday. \emph{Physics Part 1}, Third Edition (John Wiley, New York), p. 322, 323, 324.
\bibitem{foundations} J.R.Reitz, F.J.Milford, R.W.christy \emph{Foundations of electromagnetic theory}, Third Edition (Addison-wesly), p. 276, 277.
\bibitem{tout}B.Salamito, S.Cardini, D.Jurine \emph{Physique tout-en-un}, (Dunod, Paris, 2013), p. 349, 350.
\bibitem{wiki}Inversive geometry, \emph{https://$en.wikipedia.org> wiki > Inversive-geometry$}.
\bibitem{galois}Inversive geometry, \emph{https://$en.wikipedia.org> wiki > Projective-harmonic-conjugate$}.
\end{thebibliography}
\end{document}